\newcommand{\nd}{Ni(C$_5$D$_{14}$N$_2$)$_2$N$_3$(PF$_6$)}

\documentclass[aps,prl,twocolumn,superscriptaddress]{revtex4}
\usepackage{graphicx}

\usepackage{bm}

\begin{document}

\title{Massive triplet excitations in a magnetized anisotropic Haldane spin chain.}

\author{A. Zheludev}
\affiliation{Condensed Matter Sciences Division, Oak Ridge
National Laboratory, Oak Ridge, TN 37831-6393, USA.}

\author{Z. Honda}
\affiliation{Faculty of Engineering, Saitama University, Urawa,
Saitama 338-8570, Japan.}

\author{C. L. Broholm}
\affiliation{Department of Physics and Astronomy, Johns Hopkins
University, Baltimore, MD 21218, USA.}

\affiliation{NIST Center for Neutron Research, National Institute
of Standards and Technology, Gaithersburg, MD 20899, USA.}

\author{K. Katsumata}
\affiliation{The RIKEN Harima Institute, Mikazuki, Sayo, Hyogo
679-5148, Japan.}

\author{S. M. Shapiro}
\affiliation{Physics Department, Brookhaven National Laboratory,
Upton, NY 11973-5000, USA.}

\author{A. Kolezhuk}
\affiliation{Institut   f\"ur Theoretische Physik, Universit\"at
Hannover,  Appelstra{\ss}e 2, 30167, Hannover, Germany.}
\affiliation{ Institute of Magnetism, National Academy of Sciences
\& Ministry of Education of Ukraine, 36(B) Vernadskii avenue,
03142 Kiev, Ukraine.}

\author{S. Park}
\author{Y. Qiu}

\affiliation{NIST Center for Neutron Research, National Institute
of Standards and Technology, Gaithersburg, MD 20899, USA.}

\affiliation{Department of Materials and Nuclear Engineering,
University of Maryland, College Park, MD 20743, USA.}

\date{\today}
\begin{abstract}
Inelastic neutron scattering experiments on the Haldane-gap
quantum antiferromagnet \nd\ are performed at mK temperatures in
magnetic fields of almost twice  the critical field $H_c$ applied
perpendicular to the spin cahins. Above $H_c$ a re-opening of the
spin gap is clearly observed. In the high-field N\'eel-ordered
state the spectrum is dominated by three distinct long-lived
excitation branches. Several field-theoretical models are tested
in a quantitative comparison with the experimental data.
\end{abstract}


\maketitle

One-dimensional (1D) integer-spin antiferromagnets (AFs) are
famous for having a disordered ``spin liquid'' ground state and an
energy gap $\Delta\sim \exp(-\pi S)$ in the excitation spectrum
\cite{Haldane}. Elementary excitations are a triplet of massive
(gapped) long-lived ``magnons''. An external magnetic field
modifies the magnon energies by virtue of Zeeman effect
\cite{Katsumata,Affleck}. At a certain critical field $H_c$ the
gap in one of the branches approaches zero
\cite{Tsvelik90,Affleck}. The result is a condensation of
magnons\cite{Affleck,Takahashi} and the emergence of a
qualitatively new ground state. {\it What} this new ground state
actually is, depends on the symmetry of the problem. Theory
predicts that in an {\it axially symmetric} (AS) Heisenberg or
XY-like scenarios the high-field phase is a gapless ``Luttinger
spin liquid'' with quasi-long-range order and a diffuse continuum
of excitations (no sharp magnons) \cite{Sachdev94}. The high-field
phase in the {\it axially asymmetric} (AA) case is expected to be
totally different. Here the ground state should have true
long-range N\'eel order (``spin solid''). A simple boson
description \cite{Affleck} predicts a re-opening of the gap at
$H>H_c$, and implies the restoration of a single-particle
excitation spectrum. Is this state then similar to a classical
easy-plane AF in a field, that also features long-range order and
sharp gapped spin waves?

Only recently did experiments, which are the key to understanding
the high-field behavior, become technically feasible. This was in
part due to the discovery of the very useful model material \nd\
(NDMAP) \cite{Honda}, where various techniques
\cite{Honda,Zheludev01,Chen01} confirmed a quantum phase
transition at an easily accessible critical field of $H_c\approx
6$~T. Inelastic neutron studies were carried out in the AA
geometry in the {\it thermally-disordered} phase: at $H>H_c$, but
at temperatures high enough to destroy long-range N\'{e}el order
\cite{Zheludev2002}. Somewhat unexpectedly, it was found that as
the Haldane gap closes at $H_c$, the spectrum retains a
considerable quasielastic (gapless) component at higher fields.
The theoretically predicted reopening of the gap was thus not
observed. At the time, this behavior was not fully understood,
though several intriguing explanations were put forward. One
attributed the phenomenon to the 1D diffusion of thermally excited
classical solitons \cite{MikeskaSteiner}, while another drew
parallels with the incommensurate Luttinger liquid state in the AS
geometry. To better understand this issue, we carried out a new
series of measurements at considerably lower temperatures and in
higher magnetic fields, overcoming any finite-$T$ effects to
directly probe the ground state properties. The spin dynamics in
this regime was found to be qualitatively different from that
previously seen at elevated temperatures. The new data allow a
quantitative comparison with several quantum field-theoretical
models, while emphasizing dramatic differences between  the
high-field phase  and a classical magnet.

\begin{figure}
\includegraphics[width=3.3in]{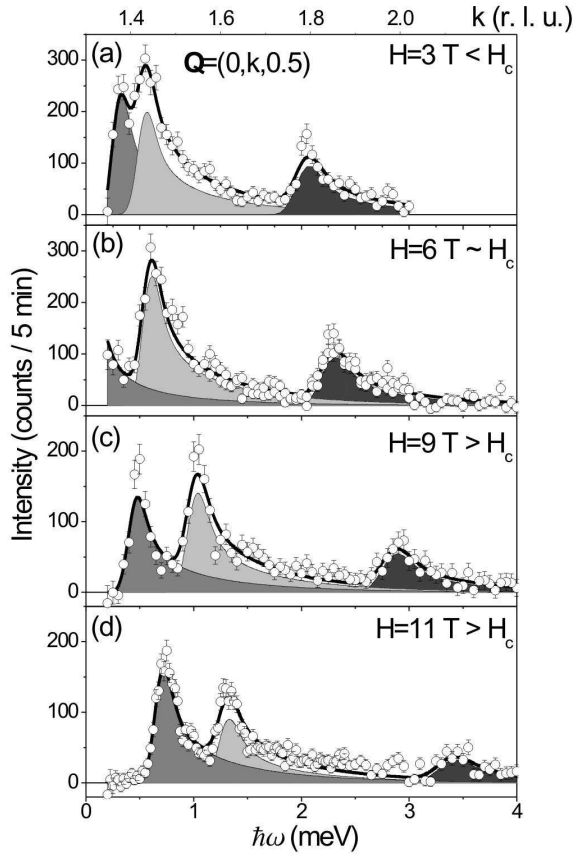}
 \caption{A series of constant-$q_\|$ scans measured in NDMAP at $T=30$~mK
 for
 different values of magnetic field applied along the $a$ axis (symbols).
 The lines are fits to a simple single-mode cross section function as described in the text.}\label{exdata}
\end{figure}

Five newly-grown deuterated NDMAP single crystals were co-aligned
by neutron diffraction to produce a sample of total mass 1.4~g and
a mosaic spread of $3^\circ$. NDMAP crystallizes
 in the orthorhombic space group $Pnmn$.
The $S=1$ AF spin chains, formed by Ni$^{2+}$ ions bridged by
azido-groups, run along the crystallographic $c$ axis, and $(a,b)$
is the magnetic easy plane. In zero field the Haldane gap energies
were previously determined to be $\Delta_x=0.42$~meV,
$\Delta_y=0.52$~meV, and $\Delta_z=1.89$~meV \cite{Zheludev01}. In
our experiments the sample was mounted with the $a$ axis vertical,
and the data were collected in the $(0,k,l)$ reciprocal-space
plane. The sample environment was a cryomagnet with a dilution
refrigerator. The measurements were performed at $T=30$~mK in
magnetic fields of up to 11~T  applied along the crystallographic
$a$ axis.

The first series of experiments was performed at the SPINS 3-axis
cold-neutron spectrometer installed at the NIST Center for High
Resolution Neutron Scattering (CHRNS). The main purpose was to
measure the field dependence of scattering at the 1D AF
zone-center $l=0.5$. Neutrons with a 3.7~meV fixed-final energy
were used with a horizontally focusing analyzer and a BeO filter
after the sample. Energy scans were performed on the $(0,k,0.5)$
reciprocal-space rod. The wave vector transfer perpendicular to
the chains was continuously adjusted to maintain the sample $c$
axis directed towards the analyzer for optimal wave vector
resolution along the chains. The background (featureless and
typically 4 counts/min) was measured away from the 1D AF
zone-center, at $(0,k,0.35)$ and $(0,k,0.65)$.

\begin{figure}
\includegraphics[width=3.3in]{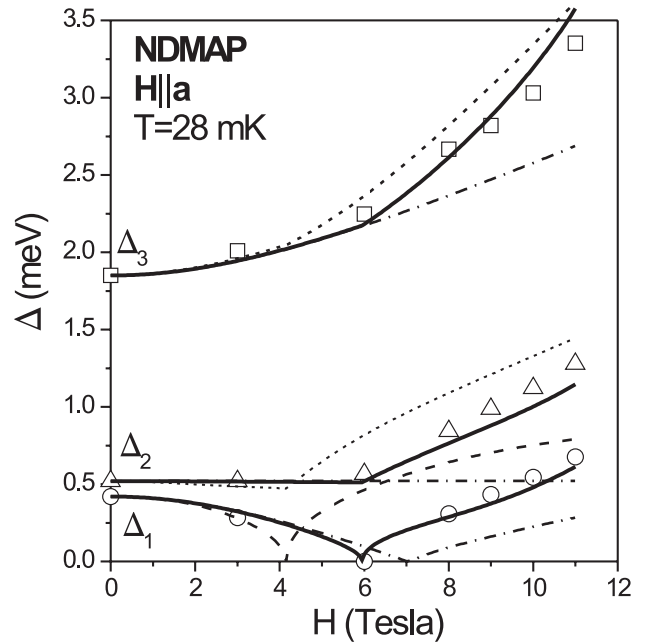}
 \caption{Measured field dependence of the gap energies in NDMAP at
 $T=30$~mK and $H$ applied along the crystallographic $a$ axis (open symbols).
 Dashed and dash-dot lines are predictions of the theoretical models
 proposed in Refs.~\protect\cite{Affleck} and \protect\cite{Tsvelik90}, respectively.
 The solid lines are a best fit to the data using an alternative
 model described in this work.}\label{results}
\end{figure}

Typical background-subtracted data sets are shown in
Fig.~\ref{exdata}. A similar scan previously measured in zero
field (Fig.~3 in Ref.~\cite{Zheludev01}) clearly shows two peaks
at roughly 0.47~mV and 1.9~meV energy transfer, respectively. The
data plotted in Fig.~\ref{exdata}a corresponds to $H=3$~T, still
well below $H_c\approx 6$~T. At this field the lower-energy peak
is visibly split in two components. At $H=6$~T$\approx H_c$ the
gap in the lower mode vanishes altogether, to within experimental
error, as illustrated in Fig.~\ref{exdata}b. At the same time,
long-range AF order sets in \cite{Chen01}. The main new result of
the present study is the observation that at $T=30$~mK at $H>H_c$
the {\it gap in the lower mode re-opens} and increases with field
(Fig.~\ref{exdata}c,d). In the ordered state the spectrum at
$q_\|=\pi$ thus contains {\it three distinct sharp excitation
branches}, just as in the low-field disordered phase.

The new 30~mK data clearly show that the quasielastic scattering
previously observed at $H>H_c$ and $T=2$~K is absent, and must
therefore be a {\it finite-$T$ effect}. In the constant-$q$ scans
collected at $T=30$~mK  all three peaks have resolution-limited
widths at all fields. In fact, very good fits to the data (solid
lines in Fig.~\ref{exdata}) can be obtained using a simple model
cross section that involves three excitations with a zero
intrinsic energy width~\cite{Zheludev01}. The cross section was
numerically convoluted with the spectrometer resolution function,
and the adjustable parameters at each field were the three gap
energies and intensity prefactors for each mode. In
Fig.~\ref{exdata} the partial contributions of the three branches
are shown as shaded areas. The field dependence of the gap
energies deduced from these fits is shown in Fig.~\ref{results} in
open symbols.
\begin{figure}
\includegraphics[width=3.2in]{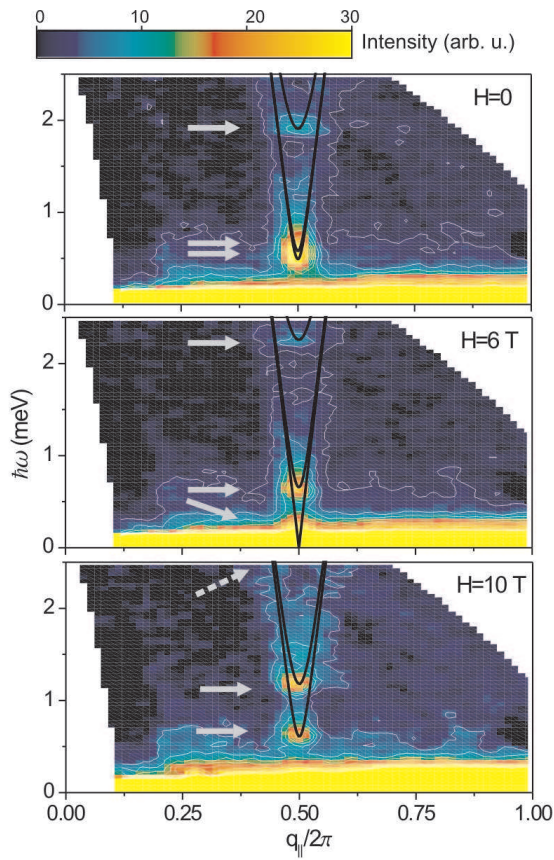}
 \caption{Inelastic spectra measured in NDMAP using the Disk Chopper spectrometer
 for several applied fields. The range of the false-color scale in the lower panel  is  0 to 15 arb. u. Contour lines are drawn with 3 arb. u. steps in all panels.
Arrows indicate the gap energies for the different excitation
branches.\label{dcs}}
\end{figure}

To study the  wave vector dependence of the dynamic structure
factor additional measurements were performed using the Disk
Chopper Spectrometer (DCS) at NIST CHRNS. The data were collected
using a fixed incident neutron energy of 4.5~meV. The sample was
mounted with the $(b,c)$ plane horizontal, and the chain axis
almost perpendicular to the incident beam. The background was
measured separately, with the sample removed from the cryostat.
The background-subtracted data collected at $H=0$, $H=6$~T and
$H=10$~T are visualized in the false-color and contour plots in
Fig.~\ref{dcs}, and correspond to a typical counting time of 20
hours. They are to be compared to similar 3-axis data measured
previously at $T=2$~K and shown in Fig.~2 of
Ref.~\cite{Zheludev2002}. The new low-$T$ experiment shows that
the excitations at $H>H_c$ have a simple relativistic (hyperbolic)
dispersion relation, with a spin wave velocity equal to that at
$H<H_c$ (see solid lines in Fig.~\ref{dcs}). The ``inverted''
hyperbolic dispersion curves with ``negative gaps'' shown in solid
lines in Fig.~3 of Ref.~\cite{Zheludev2002} are clearly
inconsistent with the new data. This latter dispersion form was
proposed as one possible interpretation of the anomalous $q$-width
of quasielastic scattering at $T=2$~K \cite{Zheludev2002} and is
based on a Fermion representation of excitations in {\it
isotropic} Haldane spin chains. The present data show that this
model does not apply in highly anisotropic case of NDMAP: the
spectrum is truly gapped and has no intrinsic incommensurate
features.  The quasielastic scattering at $T=2$~K, $H>H_c$ is thus
to be attributed to a diffusion of thermally excited topological
solitons~\cite{Zheludev2002},  and its anomalous $q$-width is due
to the $T$-dependent mean distance between solitons
\cite{MikeskaSteiner}.

For the following discussion of the observed low-$T$ properties it
is crucial to note that at $T=30$~mK the spin chains are
antiferromagnetically {\it ordered} at $H>H_c$ \cite{Chen01}, with
a static staggered magnetization as large as $m\sim
1\mu_\mathrm{B}$ per site at $H=11$~T. Clearly, inter-chain
coupling is needed to stabilize order at a non-zero temperature.
However, in the AA geometry, even an {\it isolated} chain orders
at $H>H_c$ at $T=0$, the system being equivalent to the
(1+1)-dimensional Ising-model \cite{Affleck}. Since inter-chain
interactions in NDMAP are very weak \cite{Zheludev01}, we can
assume that a purely 1D problem is realized: long-range AF
correlations are intrinsic to the 1D chains, and the sole role of
residual 3D interactions is to maintain their stability at a
finite temperature.

The conventional approach to describing spin excitations in
ordered systems is the quasiclassical spin wave theory (SWT). In
this model the magnons are {\it precessions} of staggered
magnetization $\mathbf{m}$ around its equilibrium direction. As a
consequence, in SWT there are only two sharp excitation branches,
polarized perpendicular to the $\mathbf{m}$. In our case  {\it
three} sharp magnons are seen in the ordered state ($H>H_c$). At
least one of the three branches must have the character of a
``longitudinal'' magnon that is not a precession mode, but is
polarized along the ordered moment. Thus, at $H>H_c$
quantum-mechanical effects remain crucial, and the SWT is
inapplicable. Instead, the three observed excitation branches can
be visualized as soliton-antisoliton breathers: the three massive
bound states formed by the two types of topological defects
allowed in an anisotropic semisiclassical 1D magnet
\cite{MikeskaSteiner}. In more detail our experiments can be
understood in the frameworks of several field-theoretical models.

The approach due to Affleck \cite{Affleck} is based on
coarse-graining the (1+1)-dimensional $O(3)$ non-linear sigma
model (NLSM),  to which, in turn, the $S=1$ Heisenberg chain can
be mapped \cite{Haldane}. The resulting Lagrangian
 is that of an unconstrained {\it real}
vector field $\bm{\varphi}(\mathbf{r})$ with the
$\varphi^{4}$-type interaction. Anisotropy is added by postulating
separate masses $\Delta_{\alpha}$ for the different components of
this vector field.  For $H>H_{c}$ the ground state has a non-zero
staggered magnetization $\mathbf{L}=\langle\bm{\varphi}\rangle$
and uniform magnetization $\mathbf{M}\propto
\langle\mathbf{H}\times \bm{\varphi}\rangle$. The
$\varphi^4$-model captures the basic physics involved, but suffers
from several drawbacks. In particular, the predicted value of the
critical field for $\mathbf{H}\| \mathbf{e}_{\alpha}$ is
$g\mu_BH_{c}^{(\alpha)}=\Delta_{\alpha}$. This is inconsistent
with established experimental
\cite{Regnault94,Honda97,Honda99,Chen01,Zheludev01Z} and numerical
\cite{Golinelli} results, as well as with simple arguments based
on the perturbation theory \cite{Golinelli,Regnault93}. Another
potential weakness is that at the mean field (MF) level $M(H)$ and
the lowest gap $\Delta(H)$ come out to be
$\propto\sqrt{H-H_c^{(\alpha)}}$, while a roughly linear behavior
is seen experimentally and numerically. For NDMAP, the derivations
of Ref.\ \onlinecite{Affleck} must be generalized to allow for an
arbitrary field direction, since the anisotropy axis of the ${\rm
Ni^{2+}}$ ions forms an angle of about $16^{\circ}$ with the
crystallographic $c$ axis. The resulting predictions are shown in
dashed lines in Fig.~\ref{results}.

Another theory, due to Tsvelik \cite{Tsvelik90}, stems from the
integrable  $SU(3)$ model of a $S=1$ chain. It involves three
Majorana fields with masses $\Delta_{\alpha}$. The  predicted
critical fields
$g\mu_BH_{c}^{(\alpha)}=\sqrt{\Delta_{\beta}\Delta_{\gamma}}$
coincide with the perturbative formulas of
\cite{Golinelli,Regnault93}. Again, incorporating
 an arbitrary field direction, and
using the known parameters for NDMAP, we can directly compare the
predicted gap energies (dash-dot lines in Fig.~\ref{results}) to
those measured in this work. Unlike the $\varphi^4$ model,
Tsvelik's approach predicts the linear behavior of $M(H)$ and
$\Delta(H)$ near $H_c$. The theory, however, fails to reproduce
the two upper gaps at $H>H_{c}$.

In the context of the present experiments we would like to
introduce  a different approach that is similar to the model
proposed for dimerized $S=1/2$ chains in Ref.\
\onlinecite{Kolezhuk96}. It is a Ginzburg-Landau-type theory
written in terms of a \emph{complex} triplet field
$\bm{\Phi}(\mathbf{r})=\bm{A}(\mathbf{r})+i\bm{B}(\mathbf{r})$.
The uniform and staggered magnetization are written as
$\mathbf{M}\propto (\bm{A}\times \bm{B})$ and
$\mathbf{L}\propto\bm{A}(1-A^{2}-B^{2})^{1/2}$. Interaction terms
$(\bm{\Phi}^{*}\cdot\bm{\Phi})^{2}$ and
$(\bm{\Phi}^{*}\times\bm{\Phi})^{2}$ are naturally present in the
model.  Unlike the models discussed above, anisotropy enters the
Lagrangian through \emph{two} sets of masses, separately in the
$A$ and $B$ ``channels'': $\sum_{\alpha} (m_{\alpha}
A_{\alpha}^{2}+\widetilde{m_{\alpha}} B_{\alpha}^{2})$. By
integrating out the $\bm{B}$-field one obtains an effective
$\varphi^4$-type theory similar to that of Affleck.  However, the
Zeeman term $\bm{H}\cdot (\bm{\varphi}\times
\partial_{t}\bm{\varphi})$ of Ref.\ \onlinecite{Affleck} becomes replaced by the
anisotropy-dependent expression $\sum_{\alpha\beta\gamma}
(\widetilde{m_{\gamma}})^{-1}
\varepsilon_{\alpha\beta\gamma}H_{\alpha}A_{\beta}(\partial_{t}A_{\gamma})$.
Remarkably, this model includes Affleck's theory as the special
case $\widetilde{m_{\gamma}}=\text{const}$, and at the same time
it reproduces the results of the Tsvelik's theory for $\Delta(H)$
and $M(H)$ \emph{below $H_{c}$} in another special case
$\widetilde{m_{\gamma}}=m_{\gamma}$. There are {\it no particular
reasons} why either of these special cases should correspond to
the actual Heisenberg $S=1$ chain with single-ion anisotropy
realized in NDMAP. The masses $\widetilde{m_{\alpha}}$ and
$m_{\alpha}$  at the present stage may be treated as adjustable
parameters to reproduce the measured gap energies
$\Delta_{\alpha}=\sqrt{\widetilde{m_{\alpha}}m_{\alpha}}$ at $H=0$
and the measured field dependencies. Very good fits  to our
experimental data are obtained with $\widetilde{m_x}/m_x=0.87$,
$\widetilde{m_y}/m_y=0.83$, and $\widetilde{m_z}/m_z=0.35$ (solid
lines in Fig.~\ref{results}). A detailed description of the
application of this model to NDMAP will be reported elsewhere.

In summary, the present low-$T$ study is a direct observation of
several fundamental quantum-mechanical features predicted for
anisotropic Haldane spin chains, and helps clarify the nature of
the previously investigated finite-$T$ spin dynamics. The next
experimental challenge will be to investigate the AS geometry,
searching for manifestations of the Luttinger spin liquid regime.

We would like to thank H.-J.~Mikeska,  F.~E{\ss}ler, A.~Tsvelik,
and I.~Zaliznyak for enlightening discussions. Work at ORNL and
BNL was carried out under DOE Contracts No. DE-AC05-00OR22725 and
DE-AC02-98CH10886, respectively. Work at JHU was supported by the
NSF through DMR-0074571. Experiments at NIST were supported by the
NSF through DMR-0086210 and DMR-9986442. The high-field magnet was
funded by NSF through DMR-9704257. Work at RIKEN was supported in
part by a Grant-in-Aid for Scientific Research from the Japan
Sosciety for the Promotion of Science. One of us (AK) was
supported by the grant I/75895 from Volkswagen-Stiftung.


\end{document}